\documentclass[11pt,a4paper]{article}
\usepackage{jheppub}
\usepackage{amssymb,amsmath,xcolor,graphicx,xspace,colortbl,rotating}
\usepackage{amsmath}
\usepackage{amssymb,amsmath,xcolor,graphicx,xspace,colortbl,rotating}
\usepackage{amssymb,amsmath,xcolor,graphicx,xspace,colortbl,textcomp, rotating}

\graphicspath{{Critical Log Gravity 01_graphics/}{Critical Log Gravity 01_tcache/}{Critical Log Gravity 01_gcache/}}
\DeclareGraphicsExtensions{.pdf,.eps,.ps,.png,.jpg,.jpeg}
\graphicspath{{Critical Log Gravity 01_graphics/}{Critical Log Gravity 01_tcache/}{Critical Log Gravity 01_gcache/}}
\DeclareGraphicsExtensions{.pdf,.eps,.ps,.png,.jpg,.jpeg}
\graphicspath{{Critical_gravity_graphics/}{Critical_gravity_tcache/}{Critical_gravity_gcache/}}
\DeclareGraphicsExtensions{.pdf,.eps,.ps,.png,.jpg,.jpeg}
\graphicspath{{Critical_gravity_graphics/}{Critical_gravity_tcache/}{Critical_gravity_gcache/}}
\DeclareGraphicsExtensions{.pdf,.eps,.ps,.png,.jpg,.jpeg}

\title{Holographic correlation functions in Critical Gravity }
\author[]{Giorgos Anastasiou and}
\author[]{Rodrigo Olea}

\affiliation[]{Departamento de Ciencias F{\'\i}sicas, Universidad Andres Bello, Sazi{\'e} 2212, Piso 7, Santiago, Chile }

\emailAdd{ georgios.anastasiou@unab.cl }
\emailAdd{ rodrigo.olea@unab.cl }

\abstract
{We compute the holographic stress tensor and the logarithmic energy-momentum tensor of Einstein-Weyl gravity at the critical point. This computation is carried out performing a holographic expansion in a bulk action supplemented by the Gauss-Bonnet term with a fixed coupling. The renormalization scheme defined by the addition of this topological term has the remarkable feature that all Einstein modes are identically cancelled both from the action and its variation. Thus, what remains comes from a nonvanishing Bach tensor, which accounts for non-Einstein modes associated to logarithmic terms which appear in the expansion of the metric. In particular, we compute the holographic $1$-point functions for a generic boundary geometric source.}

\begin{document}
\maketitle

\section{Introduction}

Critical Gravity belongs to a class of theories characterized
by the presence of higher curvature terms in the action.
Higher-derivative gravity theories were introduced as possible
toy models that might provide insight in some aspects of
quantum gravity. The failure of General Relativity (GR) to be
perturbatively non-renormalizable, leads to a theory that is UV
divergent, and consequently, it is not consistent at a quantum
level \cite{Goroff:1985th}.
\par
One of the proposals to solve the problem of renormalizability
was the addition of quadratic curvature terms on top of the
Einstein-Hilbert action. Seminal papers on the topic show that
these theories are renormalizable \cite{Stelle:1976gc,Adler:1982ri}.
The spectrum of the theory, for a flat spacetime, consists of
massless spin-2, massive spin-2 and massive scalar excitations.
However, as it was later pointed out, the massive graviton is a ghost mode (negative energy)
in a generic higher-derivative theory \cite{Stelle:1977ry}.
\par
In quest for consistent gravity theories, 3D massive gravity
provided intuition and the appropriate tools to overcome the
pathologies mentioned above \cite{Bergshoeff:2009hq,Deser:2002iw}.
Some  of the desirable features in these theories  can be extended
to higher  dimensions.
More specifically, the phenomenon of criticality, which represents
the existence of a point in the parametric space of the coupling
constants where the linearized EOM degenerate and the massive
gravitons turns to massless, has been extended in 4D giving rise
to Critical Gravity \cite{Lu:2011zk}. At this specific point the
scalar excitations vanish whereas new modes with logarithmic
behavior arise.
\par
The logarithmic modes can be discarded imposing standard AdS boundary
conditions. In this case, the theory is proved to be trivial as
the energy of the massless excitations as well as the energy and the
entropy of the Schwarzschild-AdS black hole vanish \cite{Lu:2011zk,Anastasiou:2017rjf}.
This is a consequence of the on-shell equivalence between Einstein-AdS
and Conformal Gravity when switching off the non-Einstein modes,
as previously seen in Refs. \cite{Maldacena:2011mk,Anastasiou:2016jix}.
\par
If, on the contrary, one imposes a relaxed set of AdS boundary
conditions, then the logarithmic modes can be included in
the spectrum of the theory. Such asymptotic conditions have
been discussed in Refs. \cite{Alishahiha:2011yb,Bergshoeff:2011ri,Gullu:2011sj}.
The behavior of the new modes is captured by terms with logarithmic dependence
in the radial coordinate in the Fefferman-Graham (FG) expansion.
The term at leading log order is the source of a logarithmic operator
living on the boundary. The boundary field theory is a Logarithmic
Conformal Field Theory (LCFT), instead of a regular CFT.
\par
In general, the presence of a logarithmic source modifies the
asymptotic structure and the spacetime fails to be asymptotically AdS
in the standard sense. As a consequence, standard holographic description
at its boundary breaks down.The alternative is treating the problem perturbatively,
with a log contribution which is very small, such that
the conformal structure at asymptotic infinity is preserved
and the holographic dictionary is still valid.
\par
LCFTs emerge in different fields in Physics, but they are mainly
associated to critical behavior of disordered systems.
Other setups where they are physically relevant include the description of
polymers, percolation, turbulence, Quantum
Hall plateau phase transition and in string theory, as well.
LCFTs are characterized by the presence of logarithmic terms in the
operator product expansion (OPE) \cite{Gurarie:1993xq} which,
despite its logarithmic behavior, respect conformal invariance.
Logarithmic operators, which in the gravity dual description
are sourced by $b_{\left(0\right)ij}$, extent the notion of a
primary operator for non-diagonalizable matrices. In particular,
they arise as logarithmic partners coupled to zero norm primary
states with degenerate scaling dimensions \cite{Caux:1995nm}.
The Hamiltonian corresponding to these states is not Hermitian,
and, therefore, they are associated to theories which are non-unitary.
\par
In view of all above arguments, it is clear that
Critical Gravity provides important intuition on properties
of the AdS/LCFT correspondence. Some interesting properties of this gravity theory
were made manifest in Ref. \cite{Anastasiou:2016jix}, where it was shown that the only
non-trivial contributions in Critical Gravity are coming from
the non-Einstein sector of the theory, as the on-shell action is quadratic
in the Bach tensor. In the present paper, we exploit this
feature in order to gain a new insight into the properties of the theory.
We identify the non-Einstein modes as the source of the divergences and
propose a new set of counterterms which depend  on the extrinsic curvature
and its covariant derivative, in order to regulate the action.
This formulation provides a shortcut in the derivation of holographic correlation functions,
as it substantially simplifies the computations respect to
similar approaches in the literature (e.g., Ref. \cite{Johansson:2012fs}).

\newpage
\section{Critical Gravity}

\noindent Critical Gravity in $4D$ is defined by the action

\begin{equation}
I_{critical} = \frac{1}{16 \pi G} \int \limits_{\mathcal{M}} d^{4}x \sqrt{-g}
\left[R - 2 \Lambda + \frac{3}{2 \Lambda} \left(R_{\mu \nu} R^{\mu \nu} -
\frac{1}{3} R^{2} \right) \right]  \label{criticalgravityaction}
\end{equation}

\noindent where $\Lambda = -3/ \ell^{2}$ is the cosmological constant (in
terms of the AdS radius $\ell$). An equivalent form for the Critical Gravity
action was provided in Refs. \cite{Miskovic:2014zja,Anastasiou:2016jix}

\begin{equation}
I_{critical} = I_{MM} - \frac{\ell^{2}}{64 \pi G} \int \limits_{\mathcal{M}%
} d^{4}x \sqrt{-g} W^{\alpha \beta \mu \nu} W_{\alpha \beta \mu \nu}
\label{criticalgravity}
\end{equation}

\noindent where $I_{MM}$ stands for
the Einstein-AdS action suitably regulated by the addition of
the Gauss-Bonnet term

\begin{equation}
I_{MM} = \frac{\ell^{2}}{256 \pi G} \int \limits_{\mathcal{M}} d^{4}x \sqrt{%
-g} \delta_{\left[\mu_{1} \mu_{2} \mu_{3} \mu_{4} \right]}^{\left[ \nu_{1}
\nu_{2} \nu_{3} \nu_{4} \right]} \left( R_{\nu_{1} \nu_{2}}^{\mu_{1}
\mu_{2}} + \frac{1}{\ell^{2}} \delta_{\left[ {\nu_{1} \nu_{2}} \right]}^{%
\left[\mu_{1} \mu_{2} \right]} \right) \left( R_{\nu_{3} \nu_{4}}^{\mu_{3}
\mu_{4}} + \frac{1}{\ell^{2}} \delta_{\left[{\nu_{3} \nu_{4}} \right]}^{%
\left[\mu_{3} \mu_{4} \right]} \right)\,.  \label{reneinsteinads}
\end{equation}

This particular form of the AdS gravity action is referred to as
MacDowell-Mansouri (Stelle-West) form in the literature \cite{MacDowell:1977jt}.
It was shown in Ref. \cite{Olea:2005gb}, that the addition of the
Gauss-Bonnet term to the Einstein-Hilbert action with negative
cosmological constant induces an extrinsic regularization scheme
for AdS gravity. It was later shown in Ref. \cite{Miskovic:2009bm},
that the use of holographic techniques in asymptotically AdS spaces
allows to expand the fields and to prove that the above action is
the renormalized action that  appears in the context of AdS/CFT
correspondence \cite{Balasubramanian:1999re,Emparan:1999pm}.

Henceforth, we shall adopt the name $I_{MM}$ for the EH plus GB
action, as we anticipate that this part of the Critical Gravity action, only
by itself, will no longer be renormalized when log terms are present.

\begin{equation}
W_{\mu \nu}^{\alpha \beta} = R_{\mu \nu}^{\alpha \beta} - \frac{1}{2}
\left(R_{\mu}^{\alpha} \delta_{\nu}^{\beta} - R_{\mu}^{\beta} \delta
_{\nu}^{\alpha} - R_{\nu}^{\alpha} \delta_{\mu}^{\beta} + R_{\nu}^{\beta}
\delta_{\mu}^{\alpha} \right) + \frac{R}{6} \delta_{\left[\mu \nu \right] }^{%
\left[\alpha \beta \right] } \,,  \label{genweyl}
\end{equation}

\noindent is the Weyl tensor of the spacetime. Here,
we will refer to the Weyl$^{2}$ part within the action of Critical Gravity
(\ref{criticalgravity}) as Conformal Gravity (CG), even though it comes with
a specific coupling

\begin{equation}  \label{Weylsquared}
I_{CG}= \frac{\ell^{2}}{256 \pi G} \delta_{\left[\mu_{1} \mu_{2} \mu_{3}
\mu_{4} \right]}^{\left[\nu_{1} \nu_{2} \nu_{3} \nu_{4} \right]} W_{\nu_{1}
\nu_{2}}^{\mu_{1} \mu_{2}} W_{\nu_{3} \nu_{4}}^{\mu_{3} \mu_{4}} \,. \\
\end{equation}

The coupling in front of the above action is such that the Einstein
modes are exactly cancelled out from Eq. (\ref{criticalgravityaction}). In other
words, as it was shown in Refs. \cite{Miskovic:2014zja,Anastasiou:2016jix},
the Critical Gravity action is identically zero for Einstein spacetimes.

As it is useful for the present derivation, we briefly review this result
below.

\subsection{Field equations}

The corresponding field equations for Critical Gravity are given by

\begin{equation}
G_{\nu}^{\mu} + \frac{\ell^{2}}{4} B_{\nu}^{\mu} = 0 \,,  \label{EOM}
\end{equation}

where $G_{\nu}^{\mu}$ is the Einstein tensor with negative cosmological
constant

\begin{equation}
G_{\nu}^{\mu} = R_{\nu}^{\mu} - \frac{1}{2} R \delta_{\nu}^{\mu} - \frac{3}{%
\ell^{2}} \delta_{\nu}^{\mu} = - \frac{1}{4} \delta_{\left[\nu \gamma \delta %
\right]}^{\left[\mu \alpha \beta \right]} \left( R_{\alpha \beta}^{\gamma
\delta} + \frac{1}{\ell^{2}} \delta_{\left[ \alpha \beta \right]}^{\left[%
\gamma \delta \right]} \right) \,\;\text{.}\;
\end{equation}

The Bach tensor $B_{\nu}^{\mu}$ is a four-derivative object that involves
the covariant derivative of the Cotton tensor and a part which is quadratic
in the curvature. It can be also written down in terms of the Weyl tensor as

\begin{equation}
B_{\nu}^{\mu} = - 4 \left( \nabla^{\alpha} \nabla_{\beta} W_{\alpha
\nu}^{\beta \mu} + \frac{1}{2} R_{\beta}^{\alpha} W_{\alpha \nu}^{\beta \mu}
\right)\;\text{.}\;
\end{equation}

The above relation makes manifest its traceless property. The fact $%
B_{\nu}^{\mu}$ is covariantly constant derives from Bianchi identity.

Taking the trace of (\ref{EOM}), we notice that the Ricci scalar does not
differ from the case of General Relativity ($R=-12/\ell^{2}$). Plugging in
the general form of the Ricci scalar into the EOM, we obtain

\begin{equation}
R_{\mu \nu} = - \frac{3}{\ell^{2}} g_{\mu \nu} - \frac{\ell^{2}}{4} B_{\mu
\nu}\,,  \label{riccicritical}
\end{equation}

what governs not only the bulk dynamics, but also determines the asymptotic
form of the boundary terms.

\subsection{On-shell action}

When Eq. (\ref{riccicritical}) is substituted in Eq. (\ref{genweyl}), one
obtains a generic decomposition of the Weyl tensor into an Einstein and a
non-Einstein parts

\begin{equation}
W_{\mu \nu}^{\alpha \beta} = W_{\left(E\right) \mu \nu}^{\alpha \beta} +
W_{\left(NE\right) \mu \nu}^{\alpha \beta}\,,  \label{weyldecomp}
\end{equation}

where

\begin{eqnarray}
W_{\left(E\right) \mu \nu}^{\alpha \beta} &=& R_{\mu \nu}^{\alpha \beta} +
\frac{1}{\ell^{2}} \delta_{\left[\mu \nu \right]}^{\left[\alpha \beta \right]%
} \,,  \label{EWeyl} \\
W_{\left(NE\right) \mu \nu}^{\alpha \beta} &=& \frac{\ell^{2}}{8}
\left(B_{\mu}^{\alpha} \delta_{\nu}^{\beta} - B_{\mu}^{\beta}
\delta_{\nu}^{\alpha} - B_{\nu}^{\alpha} \delta_{\mu }^{\beta} +
B_{\nu}^{\beta} \delta_{\mu}^{\alpha} \right) \,.  \label{NEWeyl}
\end{eqnarray}

Here $W_{\left(E\right) \mu \nu}^{\alpha \beta}$ corresponds to the Weyl
tensor for Einstein spacetimes $R_{\mu \nu} = - 3/\ell^{2} g_{\mu \nu}$. The
departure from the Einstein condition provides additional contributions to
the Weyl tensor. In Einstein-Weyl gravity, the deviation from Einstein
spaces involves linear terms in the Bach tensor.

\newpage
Applying the Weyl decomposition (\ref{weyldecomp}) in the Weyl$^2$ part
of the action (\ref{criticalgravity}) leads to the expression

\small
\begin{equation}  \label{Weyl2}
I_{CG} =  \frac{\ell^{2}}{256 \pi G} \delta_{\left[\mu_{1} \mu_{2} \mu_{3}
\mu_{4} \right]}^{\left[\nu_{1} \nu_{2} \nu_{3} \nu_{4} \right]}
\left(W_{\left(E\right) \nu_{1} \nu_{2}}^{\mu_{1} \mu_{2}} W_{\left(E\right)
\nu_{3} \nu_{4}}^{\mu_{3} \mu_{4}} + 2 W_{\left(E\right) \nu_{1}
\nu_{2}}^{\mu_{1} \mu_{2}} W_{\left(NE\right) \nu_{3} \nu_{4}}^{\mu_{3}
\mu_{4}} + W_{\left(NE\right) \nu_{1} \nu_{2}}^{\mu_{1} \mu_{2}}
W_{\left(NE\right) \nu_{3} \nu_{4}}^{\mu_{3} \mu_{4}} \right) \,.
\end{equation}

\normalsize

The first term in the above expression carries a particular
coupling constant, such that it exactly cancels the $I_{MM}$ part in the
Critical Gravity action (\ref{criticalgravity}). This is a direct
consequence of the equivalence between Conformal and Einstein gravity, once
Neumann boundary conditions are imposed in order to get rid of
higher-derivative modes \cite{Maldacena:2011mk}. An explicit proof of this
statement, carried out in Ref. \cite{Anastasiou:2016jix}, recovers Einstein
gravity by imposing $B_{\mu \nu}=0$ in the decomposition of the $Weyl^{2}$
term (\ref{Weyl2}).\newline

As a consequence, the only nonvanishing part of Critical Gravity action is
given in terms of the Bach tensor

\begin{equation}
I_{critical} = - \frac{\ell^{4}}{64 \pi G} \int \limits_{\mathcal{M}} d^{4}x
\sqrt{-g} \delta_{\left[\mu \nu \right]}^{\left[\kappa \lambda \right]}
\left( \frac{\ell^{2}}{8} B_{\kappa}^{\mu} + G_{\kappa}^{\mu} \right)
B_{\lambda}^{\nu} \,.
\end{equation}

Using the equation of motion (\ref{EOM}),

\begin{eqnarray}
I_{critical} &=& \frac{\ell^{6}}{512 \pi G} \int \limits_{\mathcal{M}}
d^{4}x \sqrt{-g} \delta_{\left[\mu \nu \right]}^{\left[\kappa \lambda \right]%
} B_{\kappa}^{\mu} B_{\lambda}^{\nu}  \notag \\
&=& - \frac{\ell^{6}}{512 \pi G} \int \limits_{\mathcal{M}} d^{4}x \sqrt{-g}
B_{\kappa}^{\mu} B_{\mu}^{\kappa} \,,  \label{CGonshell}
\end{eqnarray}

one can notice that Critical Gravity action involves only the non-Einstein
part of the Weyl tensor in the form of the square of the Bach tensor.

\subsection{Surface terms}

An arbitrary variation of the action (\ref{reneinsteinads}) is
given by

\begin{equation}
\delta I_{MM} = \frac{\ell^{2}}{64 \pi G} \int \limits_{ \partial \mathcal{M%
}} d^{3}x \sqrt{-h} \delta_{\left [\mu_{1} \mu_{2} \mu_{3} \mu_{4} \right]}^{%
\left[\nu_{1} \nu_{2} \nu_{3} \nu_{4} \right]} n_{\nu_{1}} \delta
\Gamma_{\kappa \nu_{2}}^{\mu_{1}} g^{\mu_{2} \kappa} W_{\left(E\right)
\nu_{3} \nu_{4}}^{\mu_{3} \mu_{4}} \,.  \label{variationeinstein}
\end{equation}

Similarly, the surface terms coming from the variation of the Weyl$^2$ term
are cast into the form

\begin{equation}
\delta I_{CG} = \frac{\ell ^{2}}{64 \pi G} \int \limits_{ \partial
\mathcal{M}} d^{3}x \sqrt{-h} \delta_{\left[\mu_{1} \mu_{2} \mu_{3} \mu_{4} %
\right]}^{\left[\nu_{1} \nu_{2} \nu_{3} \nu_{4} \right]} \left[n_{\nu_{1}}
\delta \Gamma_{\kappa \nu_{2}}^{\mu_{1}} g^{\mu_{2} \kappa} W_{\nu_{3}
\nu_{4}}^{\mu_{3} \mu_{4}} + n^{\mu_{1}} \nabla_{\nu_{1}} W_{\nu_{2}
\nu_{3}}^{\mu_{2} \mu_{3}} \left(g^{-1} \delta g \right)_{\nu_{4}}^{\mu_{4}} %
\right] \nonumber \,.
\label{variationweylsquared}
\end{equation}

\newpage
Combining these two contributions, we get the total surface term of Critical
Gravity action

\begin{eqnarray}
\delta I_{critical} = \delta I_{MM} - \delta I_{CG} &=& \frac{\ell^{2}}{%
64 \pi G} \int \limits_{ \partial \mathcal{M}} d^{3}x \sqrt{-h} \delta_{%
\left[\mu_{1} \mu_{2} \mu_{3} \mu_{4} \right]}^{\left[\nu_{1} \nu_{2}
\nu_{3} \nu_{4} \right]} \left[ n_{\nu_{1}} \delta \Gamma_{\kappa
\nu_{2}}^{\mu_{1}} g^{\mu_{2} \kappa} \left(W_{\left(E\right) \nu_{3}
\nu_{4}}^{\mu_{3} \mu_{4}} - W_{\nu_{3} \nu_{4}}^{\mu_{3} \mu_{4}} \right)
\right.  \notag \\
&-& \left. n^{\mu_{1}} \nabla_{\nu_{1}} W_{\nu_{2} \nu_{3}}^{\mu_{2}
\mu_{3}} \left(g^{-1} \delta g \right)_{\nu_{4}}^{\mu_{4}} \right] \,.
\label{generalSTcriticalgravity}
\end{eqnarray}

Applying the Weyl decomposition (\ref{weyldecomp}) and the Bianchi identity
in the previous expression, one gets

\begin{eqnarray}
\delta I_{critical} &=& - \frac{\ell^{2}}{64 \pi G} \int \limits_{ \partial
\mathcal{M}} d^{3}x \sqrt{-h} \delta _{\left[\mu_{1} \mu_{2} \mu_{3} \mu_{4}
\right]}^{\left[\nu_{1} \nu_{2} \nu_{3} \nu_{4} \right]} \nonumber \\ 
& & \times \left[n_{\nu_{1}} \delta \Gamma_{\kappa \nu_{2}}^{\mu_{1}} g^{\mu_{2} \kappa}
W_{\left(NE\right) \nu_{3} \nu_{4}}^{\mu_{3} \mu_{4}} + n^{\mu_{1}}
\nabla_{\nu_{1}} W_{(NE) \nu_{2} \nu_{3}}^{\mu_{2} \mu_{3}} \left(g^{-1}
\delta g \right)_{\nu_{4}}^{\mu_{4}} \right] \,.
\label{STcriticalgravitydecomp}
\end{eqnarray}

\normalsize

In the last step, the Bianchi identity is applied as follows

\begin{equation}
\delta_{\left[\mu_{1} \mu_{2} \mu_{3} \mu_{4} \right]}^{\left[\nu_{1}
\nu_{2} \nu_{3} \nu_{4} \right]} \nabla_{\nu_{1}} W_{\left(E\right) \nu_{2}
\nu_{3}}^{\mu_{2} \mu_{3}} = 0 \, .  \notag
\end{equation}

In order to reveal that the variation of the Critical Gravity action is
linear to the Bach tensor, we substitute the expression (\ref{NEWeyl})
in Eq. (\ref{STcriticalgravitydecomp}), what leads to

\small
\begin{equation}
\delta I_{critical} = - \frac{\ell^{4}}{128 \pi G} \int \limits_{ \partial
\mathcal{M}} d^{3}x \sqrt{-h} \delta_{\left[\mu_{1} \mu_{2} \mu_{3} \right]%
}^{\left[\nu_{1} \nu_{2} \nu_{3} \right]} \left[n_{\nu_{1}} \delta
\Gamma_{\kappa \nu_{2}}^{\mu_{1}} g^{\mu_{2} \kappa} B_{\nu_{3}}^{\mu_{3}}
+n^{\mu _{1}} \nabla_{\nu_{1}} B_{\nu_{2}}^{\mu_{2}} \left(g^{-1} \delta g
\right)_{\nu_{3}}^{\mu_{3}} \right] \,.  \label{STcriticalgravityfinal}
\end{equation}

\normalsize
A direct consequence of the above formula is the
vanishing of the energy for Einstein spacetimes. This has been pointed out
in, e.g., in Refs. \cite{Lu:2011zk,Lu:2011ks,Porrati:2011ku}, based on a rather
case-by-case analysis. A more general proof that Einstein spacetimes have zero
energy can be made by using Noether-Wald charges \cite{Anastasiou:2017rjf}.

We can replace the Bach with the Einstein tensor using the EOM (\ref{EOM}), such that

\small
\begin{equation}
\delta I_{critical} = \frac{\ell^{2}}{32 \pi G} \int \limits_{ \partial
\mathcal{M}} d^{3}x \sqrt{-h} \delta_{\left[\mu_{1} \mu_{2} \mu_{3} \right]%
}^{\left[\nu_{1} \nu_{2} \nu_{3} \right]} \left[n_{\nu_{1}} \delta
\Gamma_{\kappa \nu_{2}}^{\mu_{1}} g^{\kappa \mu_{2}} G_{\nu_{3}}^{\mu_{3}} +
n^{\mu_{1}} \nabla_{\nu_{1}} G_{\nu_{2}}^{\mu_{2}} \left(g^{-1} \delta g
\right)_{\nu_{3}}^{\mu_{3}} \right]  \,.
\label{ST2}
\end{equation}

\normalsize

Notice that some of the terms of the second part of Eq.(\ref{ST2}) will vanish
due to the Bianchi identity, once the antisymmetric Kronecker delta is expanded.
Equipped with the generic form of the variation of the action (\ref{ST2}),
a suitable intermediate step towards the derivation of the holographic correlation
functions is to cast the corresponding surface terms in Gaussian coordinates,

\begin{equation}
ds^{2} = N^{2} \left(\rho \right) d \rho^{2} + h_{ij} \left(\rho,x\right)
dx^{i} dx^{j} \,.
\label{Gaussnormal}
\end{equation}

This far, Greek letters represent spacetime indices. In what follows,  Latin will denote
letters boundary indices. In this frame, the first part of Eq. (\ref{ST2}) becomes

\begin{equation}
\delta_{\left[\mu_{1} \mu_{2} \mu_{3} \right]}^{\left[\nu_{1} \nu_{2}
\nu_{3} \right]} n_{\nu_{1}} \delta \Gamma_{\kappa \nu_{2}}^{\mu_{1}}
g^{\kappa \mu_{2}} G_{\nu_{3}}^{\mu_{3}} = N \delta_{\left[k \ell \right]}^{%
\left[ij \right]} \left[\delta \Gamma_{m i}^{\rho} h^{mk} G_{j}^{\ell} -
\delta \Gamma_{\rho i}^{k} g^{\rho \rho} G_{j}^{\ell} + \delta
\Gamma_{mi}^{k} h^{m \ell} G_{j}^{\rho} \right] \,.  \label{firstST}
\end{equation}

The first two terms of Eq. (\ref{firstST}) are of the form

\begin{equation}
N \delta_{\left[k \ell \right]}^{\left[ij \right]} \left[\delta \left(\frac{1%
}{N} K_{mi} \right) h^{km} G_{j}^{\ell} - \delta \Gamma_{\rho i}^{k} g^{\rho
\rho} G_{j}^{\ell} \right] = \delta_{\left[k \ell \right]}^{\left[ij \right]%
} \left[K_{i}^{m} \left(h^{-1} \delta h \right)_{m}^{k} + 2 \delta K_{i}^{k} %
\right] G_{j}^{\ell} \,.  \label{radialST1}
\end{equation}

Moreover, the last term of Eq. (\ref{firstST}) can be written as

\begin{equation}
\int \limits_{\partial \mathcal{M}} d^{3}x \sqrt{-h} \delta_{\left[k \ell %
\right]}^{\left[ij \right]} N \delta \Gamma_{mi}^{k} h^{\ell m} G_{j}^{\rho}
= - \int \limits_{\partial \mathcal{M}} d^{3}x \sqrt{-h} N \delta_{\left[k
\ell \right]}^{\left[ij \right]} \left(h^{-1} \delta h \right)_{i}^{k}
D^{\ell} G_{j}^{\rho} \,,  \label{radialST2}
\end{equation}

where integration by parts was performed. Here, $D_{i}$ is the covariant
derivative defined in the boundary metric.

Summing up the contributions from Eqs. (\ref{radialST1}) and (\ref{radialST2}%
), one shows that Eq. (\ref{firstST}) adopts the form

\small
\begin{equation}
\delta_{\left[\mu_{1} \mu_{2} \mu_{3} \right]}^{\left[\nu _{1} \nu _{2} \nu
_{3} \right]} n_{\nu_{1}} \delta \Gamma_{\kappa \nu_{2}}^{\mu_{1}} g^{\kappa
\mu _{2}} G_{\nu_{3}}^{\mu_{3}} = \delta_{\left[k \ell \right]}^{\left[ij %
\right]} \left[\left( K_{i}^{m} \left( h^{-1} \delta h \right)_{m}^{k} + 2
\delta K_{i}^{k} \right) G_{j}^{\ell} - N D^{k} G_{i}^{\rho} \left( h^{-1}
\delta h \right)_{j}^{\ell} \right] \,.  \label{firstSTfinal}
\end{equation}

\normalsize

In addition to this, the remaining contribution coming from Eq. (\ref{ST2})
adopts the form

\begin{equation}
\delta_{\left [\mu_{1} \mu_{2} \mu_{3} \right]}^{\left[\nu_{1} \nu_{2}
\nu_{3} \right]} n^{\mu_{1}} \nabla_{\nu_{1}} G_{\nu_{2}}^{\mu_{2}}
\left(g^{-1} \delta g \right)_{\nu_{3}}^{\mu_{3}} = \delta_{\left[k \ell %
\right]}^{\left[i j \right]} \frac{1}{N} \left( \nabla _{\rho } G_{i}^{k} -
\nabla_{i} G_{\rho}^{k} \right) \left(h^{-1} \delta h \right)_{j}^{\ell} \,.
\label{secondSTfinal}
\end{equation}

Hence, the variation of the Critical Gravity action in Gauss-normal
coordinates becomes

\begin{eqnarray}
\delta I_{critical} &=& \frac{\ell^{2}}{32 \pi G} \int \limits_{\partial
\mathcal{M}} d^{3}x \sqrt{-h} \delta_{\left[k \ell \right]}^{\left[ij \right]%
} \left[ \left( 2 \delta K_{i}^{k} + K_{i}^{m} \left(h^{-1} \delta h
\right)_{m}^{k} \right) G_{j}^{\ell} \right.  \notag \\
&+& \left. \frac{1}{N} \left( \nabla_{\rho} G_{i}^{k} - \nabla_{i}
G_{\rho}^{k} - N^{2} D^{k} G_{i}^{\rho}\right) \left(h^{-1} \delta h
\right)_{j}^{\ell} \right] \,.
\label{totvaraction}
\end{eqnarray}

\section{Holographic Renormalization in Critical Gravity}

In Critical Gravity, new modes appear as a consequence of the coalescence of
massive spin-2 modes with the massless ones. These modes have logarithmic
dependence in the radial coordinate and spoil the standard asymptotically
AdS (AAdS) fall-off of the spacetime. Choosing suitable boundary conditions,
the logarithmic modes can be discarded. Thus, one reproduces the standard
AdS/CFT dictionary, sourced by Einstein modes at the boundary.

By keeping the logarithmic modes, one gains intuition on holographic duals
to higher-derivative gravity theories at critical points. In particular,
we focus on aspects of AdS/LCFT correspondence associated to the computation
of the holographic stress tensors which are defined as the functional derivatives
of the variation of the action with respect to the independent sources \cite{Witten:1998qj}.

The new branch of solutions is consistent with a relaxed set of AdS boundary
conditions \cite{Alishahiha:2011yb,Grumiller:2008es}, which is expressed in
the radial foliation (\ref{Gaussnormal}) setting $N = \frac{\ell}{2 \rho}$ and

\small
\begin{eqnarray}
& & h_{ij} \left(\rho,x \right) = \frac{1}{\rho} \tilde{g}_{ij} \left(\rho,x
\right) \,,  \label{boundarymetric} \\
& & \tilde{g}_{ij} \left (\rho,x\right) = g_{\left(0\right)ij} +
b_{\left(0\right)ij} \log \rho + \rho \left(g_{\left(2\right) ij} +
b_{\left(2\right)ij} \log \rho \right) + \rho^{3/2} \left(
g_{\left(3\right)ij} + b_{\left(3\right) ij} \log \rho \right) + ...
\label{FGlog}
\end{eqnarray}

\normalsize

\subsection{Generic boundary geometry}

In the treatment below, for simplicity, we choose unit AdS radius ($\ell = 1$).
The inverse of the boundary metric reads

\begin{eqnarray}
\tilde{g}^{ij} \left(\rho,x\right) &=& g_{\left(0\right)}^{ij} -
b_{\left(0\right)}^{ij} \log {\rho} + \rho \left [ - g_{\left(2\right)}^{ij}
- b_{\left(2\right)}^{ij} \log {\rho} + 2 \left (b_{\left(0\right)}
g_{\left(2\right)} \right)^{ij} \log {\rho} + 2 \left(b_{\left(0\right)}
b_{\left(2\right)}\right)^{ij} \log^{2} {\rho} \right]  \notag \\
&+& \rho^{3/2} \left[ - g_{\left(3\right)}^{ij} - b_{\left(3\right)}^{ij}
\log {\rho} + 2 \left(b_{\left(0\right)} g_{\left(3\right)}\right )^{ij}
\log {\rho} + 2 \left(b_{\left(0\right)} b_{\left(3\right)}\right)^{ij}
\log^{2} {\rho} \right]+ ...
\end{eqnarray}

By definition, the extrinsic curvature is given by $K_{ij} = - \frac{1}{2N} \partial_{\rho} h_{ij}$
what, in the above frame is expressed as

\begin{eqnarray}
K_{j}^{i} &=& \delta_{j}^{i} - b_{\left(0\right) j}^{i} + \rho \left[ -
b_{\left(2\right) j}^{i} - g_{\left(2\right) j}^{i} - b_{\left (2\right)
j}^{i} \log{\rho} + \left(b_{\left (0\right) } g_{\left(2\right)}
\right)_{j}^{i} + 2 \left(b_{\left(0\right) } b_{\left (2\right)}
\right)_{j}^{i} \log{\rho} + \left(b_{\left (0\right ) } g_{\left (2\right
)} \right)_{j}^{i} \log{\rho} \right.  \notag \\
&+& \left. \left(b_{\left(0\right)} b_{\left(2\right)} \right)_{j}^{i}
\log^{2}{\rho} \right] + \rho^{3/2} \left[- b_{\left(3\right) j}^{i} - \frac{%
3}{2} g_{\left (3\right ) j}^{i} - \frac{3}{2} b_{\left (3\right ) j}^{i}
\log{\rho} + \left(b_{\left (0\right ) } g_{\left (3\right)} \right)_{j}^{i}
+ 2 \left(b_{\left(0\right) } b_{\left(3\right)} \right)_{j}^{i} \log{\rho}
\right.  \notag \\
&+& \left. \frac{3}{2} \left(b_{\left(0\right) } g_{\left(3\right)}
\right)_{j}^{i} \log{\rho} + \frac{3}{2} \left(b_{\left(0\right) }
b_{\left(3\right)} \right)_{j}^{i} \log^{2}{\rho} \right] + ...
\label{extrinsiccurv}
\end{eqnarray}

As it can be easily seen from the asymptotic expansion of the extrinsic
curvature (\ref{extrinsiccurv}), the log term prevents $\delta K_{j}^{i}$
from vanishing at leading (finite) order.
Thus, on top of the boundary metric $g_{\left(0\right) ij}$, which is the
source of the boundary stress-energy tensor, a new independent source
arises, $b_{\left(0\right) ij}$. Furthermore, the presence of the new source
modifies the asymptotic expansion of the curvature as follows,

\begin{eqnarray}
& & R_{j \rho}^{i \rho} = - \delta_{j}^{i} + 2 b_{\left(0\right)j}^{i} +
\mathcal{O} \left(\rho\right)  \notag \\
& & R_{jk}^{i \rho} = 2 \rho \left( D_{k} b_{\left(0\right)j}^{i} -
D_{j} b_{\left(0\right)k}^{i} \right) + \mathcal{O} \left(\rho^{2}%
\right)  \notag \\
& & R_{kl}^{ij} = - \delta_{\left[kl\right]}^{\left[ij\right]} +
b_{\left(0\right)k}^{i} \delta_{l}^{j} - b_{\left(0\right)l}^{i}
\delta_{k}^{j} - b_{\left(0\right)k}^{j} \delta_{l}^{i} +
b_{\left(0\right)l}^{j} \delta_{k}^{i} + \mathcal{O} \left(\rho\right) \,.
\notag
\end{eqnarray}

Thus, the spacetime is no longer AAdS and as a result the dual CFT
description is not valid anymore. Nevertheless, considering a non-vanishing
but sufficiently small $b_{\left(0\right) ij}$, one avoids spoiling the
asymptotic conformal structure of the AAdS spacetime. For this reason, in
the present section, we proceed perturbatively in $b_{\left(0\right)ij}$. The
dual theory living on the boundary is now a Logarithmic Conformal Field
Theory (LCFT), instead of a CFT. As a result, a new operator arises,
which is identified as the logarithmic stress energy tensor $t_{ij}$, i.e.,
the response to $b_{\left(0\right) ij}$. In the dual description, the
logarithmic stress tensor is an irrelevant operator.

The Eq.(\ref{totvaraction}), evaluated in the FG expansion for relaxed
AdS boundary conditions (\ref{FGlog}), provides the holographic one-point
functions of the boundary field theory. These are expressed as the
coefficients of $\delta g_{\left(0\right) ij}$ and $\delta b_{\left(0\right)
ij}$ , which are the sources of the dual stress tensors. This derivation requires the
asymptotic resolution of the EOM, by substituting
the expression (\ref{FGlog}) in (\ref{EOM}). As a result, one obtains
algebraic equations relating the coefficients in FG expansion of the metric.

\subsection{Vanishing log source $(b_{\left(0\right) ij}=0)$}
As a warmup computation, here we derive the energy-momentum tensor for a
given boundary background metric $g_{\left(0\right) ij}$, while setting the
leading logarithmic mode $b_{\left(0\right) ij}$ as zero.

Thus, the trace of the equation of motion ($R = - 12$) gives rise to

\begin{align}
Tr g_{\left(3\right)} = Tr b_{\left(3\right)} = 0 = Tr b_{\left(2\right)} \,,
\\
4 Tr g_{\left(2\right)} + R \left(g_{\left(0\right)}\right) = 0 \,.
\end{align}

For the $\left(\rho i \right)$ component of the EOM one finds

\begin{align}
\nabla_{j} b_{\left(3\right) i}^{j} = \nabla_{j} g_{\left(3\right) i}^{j} =
\nabla_{j} b_{\left(2\right) i}^{j} = 0 \,, \\
\nabla_{i} Tr g_{\left(2\right)} - \nabla_{j} g_{\left(2\right) i}^{j} = 0
\,.
\end{align}

Finally, for the $\left(ij\right)$ part of the EOM

\begin{align}
g_{\left(2\right) j}^{i} - Tr g_{\left(2\right)} \delta_{j}^{i} + R_{j}^{i}
\left(g_{\left(0\right)}\right) - \frac{1}{2} R
\left(g_{\left(0\right)}\right) \delta_{j}^{i} = 0 \,, \\
b_{\left(2\right) j}^{i} = 0 \,.
\end{align}

Thus, the vanishing of $b_{\left(0\right)}$ leads to a vanishing $%
b_{\left(2\right)}$ in (\ref{FGlog}), but there is yet a remaining
logarithmic contribution coming from the subdominant $b_{\left(3\right)}$
term. Due to the absence of logarithmic source, the corresponding
energy-momentum tensor $t_{ij}$ is zero.

In this case the extrinsic curvature is expanded asymptotically as

\begin{equation}
K_{j}^{i} = \delta_{j}^{i} - \rho g_{\left(2\right) j}^{i} + \rho^{3/2}
\left(b_{\left(3\right) j}^{i} - \frac{3}{2} g_{\left(3\right) j}^{i} -
\frac{3}{2} b_{\left(3\right) j}^{i} \log \rho \right)+ ...
\label{extrcurvb0}
\end{equation}

\newpage
Calculating the contributions appearing in the action (\ref{totvaraction}),
we obtain

\begin{eqnarray}
\delta_{\left[k \ell \right]}^{\left[ij \right]} K_{i}^{m} G_{j}^{\ell} &=&
3 \rho^{3/2} b_{\left(3\right) k}^{m} \\
\frac{1}{N} \delta_{\left[k \ell \right]}^{\left[ij \right]} \left(
\nabla_{\rho} G_{i}^{k} - \nabla_{i} G_{\rho}^{k} \right) &=& 6 \rho ^{3/2}
b_{\left(3\right) \ell}^{j} \\
N \delta_{\left[k \ell \right]}^{\left[ij \right]} D^{k} G_{i}^{\rho} &=& 0
\end{eqnarray}

Moreover, the absence of $b_{\left (0\right )i j}$ turns the variation of
the extrinsic curvature into a term of order $\mathcal{O} \left(\rho \right)$%
. Considering the corresponding asymptotic expansion of the fields

\begin{eqnarray}
& & \left (h^{ -1} \delta h\right )_{j}^{i} = \left (\tilde{g}^{ -1} \delta
\tilde{g}\right )_{j}^{i} = \left (g_{\left (0\right )}^{ -1} \delta
g_{\left (0\right )}\right )_{j}^{i} + \mathcal{O} \left (\rho \right ) \\
& & \sqrt{ -h} = \frac{\sqrt{ -\tilde{g}}}{\rho ^{3/2}} = \frac{\sqrt{
-g_{\left (0\right )}}}{\rho ^{3/2}} + \mathcal{O} \left ( \rho ^{ -1/2}
\right ) \,,
\end{eqnarray}

the variation of $K_{j}^{i}$ is subdominant with respect to the variation of the metric,
as in standard AAdS spacetimes. Therefore, this terms does not contribute at the conformal boundary.
Thus, Eq. (\ref{totvaraction}) adopts the form

\begin{equation}
\delta I_{critical} =\frac{9}{32 \pi G} \int \limits_{\partial
\mathcal{M}} d^{3}x \sqrt{- g_{\left(0\right)}} b_{\left (3\right )
j}^{i} \left (g_{\left (0\right )}^{ -1} \delta g_{\left (0\right )}\right
)_{i}^{j} \,.  \label{varactTij}
\end{equation}

No infrared divergences appear and the variational principle is well
defined for the Dirichlet boundary condition, $\delta g_{\left (0\right )ij}
= 0$. Hence, it turns out that no counterterms are needed on top of
the Critical Gravity action (\ref{criticalgravityaction}).

The holographic stress tensor is obtained as the functional variation
of the regular part of the surface term respect to the metric source \cite{Witten:1998qj}

\begin{equation}
\langle T_{i j} \rangle = -\frac{2}{\sqrt{ -g_{\left (0\right )}}} \frac{%
\delta I}{\delta g_{\left (0\right )}^{i j}}\,.
\end{equation}

Reading off from the formula (\ref{varactTij}), one gets

\begin{equation}
\langle T_{i j} \rangle = \frac{9}{16 \pi G} b_{\left (3\right )i j} \,,
\end{equation}

for the holographic one-point function dual to the boundary metric $g_{\left
(0\right ) i j}$. This formula recovers the result in Ref. \cite%
{Johansson:2012fs} without assuming any particular form on the boundary
geometry.

It is clear that, for Einstein spaces, the above stress tensor is zero.

\section{Linearized analysis ($b_{\left (0\right )ij}\neq0$)}

The calculation of the holographic correlation functions, when $b_{\left (0\right )ij}$
is switched on, turns considerably cumbersome. In order to simplify the discussion,
one tackles the problem perturbatively around AdS$_{4}$ \cite%
{Alishahiha:2010bw,Johansson:2012fs,Skenderis:2009nt}. Linearizing the EOM,
one gets the on-shell action up to quadratic order in the perturbation. This
analysis is sufficient for the derivation of the two-point functions. In
this section, taking advantage of the alternative form of the Critical Gravity action (%
\ref{CGonshell}), we evaluate the linearized AdS$_{4}$ metric and introduce
proper counterterms in order to cancel the emerging infinities.

Following the analysis of the generic case in Gauss-normal
coordinates (\ref{Gaussnormal}), where the boundary metric is expressed as a
deviation of the Minkowski background,

\begin{equation}
h_{ij} \left(\rho,x\right) = \frac{1}{\rho} \tilde{g}_{ij} = \frac{1}{\rho}
\left(\eta_{ij} + c_{ij}\right) \,.  \label{linGauss}
\end{equation}

In this gauge, the holographic correlation functions are planar. The
perturbation $c_{ij}$ admits the FG expansion

\begin{equation}
c_{ij} = h_{\left(0\right) ij} + b_{\left(0\right) ij} \log \rho + \rho
\left(g_{\left(2\right) ij} + b_{\left(2\right) ij} \log \rho \right) +
\rho^{3/2} \left(g_{\left(3\right) ij} + b_{\left(3\right) ij} \log \rho
\right) + ... \,,  \label{FGperturbation}
\end{equation}

which is consistent within the logarithmic branch of the theory. The
substitution of the FG expansion (\ref{FGperturbation}) in Eq. (\ref{EOM}),
provides relations between the FG coefficients, which hold for the linearized
version of the theory.

From the Ricci scalar $R = -12$ , one is able to obtain

\begin{eqnarray}
& & Tr b_{\left (0\right )} = Tr b_{\left (3\right )}= Tr g_{\left (3\right
)} = 0  \label{trb0} \\
& & 4 Tr b_{\left (2\right )} + \partial_{i} \partial_{j} b_{\left (0\right
)}^{i j} = 0  \label{trb2} \\
& & 4 Tr g_{\left (2\right )} + \partial_{i} \partial_{j} h_{\left (0\right
)}^{i j} - \partial^{m} \partial_{m} Tr h_{\left (0\right )} = 0 \,.
\label{trg2}
\end{eqnarray}

The $\left (\rho \rho \right )$ component of the EOM (\ref{EOM}) give

\begin{equation}
4 Tr b_{\left (2\right )} - \partial_{i} \partial_{j} b_{\left (0\right
)}^{i j} = 0 \,,
\end{equation}

while the tracelessness of the Bach tensor leads to

\begin{equation}
2 Tr b_{\left (2\right )} - \frac{7}{2} \partial_{i} \partial_{j} b_{\left
(0\right )}^{i j} = 0 \,.
\end{equation}

Combining these expressions with Eq. (\ref{trb2}), one concludes that

\begin{equation}
Tr b_{\left (2\right )} = \partial_{i} \partial_{j} b_{\left (0\right )}^{i
j} = 0 \,.
\end{equation}

The $\left (\rho i\right )$ terms give

\begin{eqnarray}
& & \partial_{j} b_{\left (0\right ) i}^{j} = \partial_{j} b_{\left (2\right
) i}^{j} = \partial_{j} b_{\left (3\right ) i}^{j} = 0 \,, \label{divb0} \\
& & 4 \partial_{j} g_{\left (2\right ) i}^{j} + \partial_{i} \partial_{m}
\partial_{k} h_{\left (0\right )}^{m k} - \partial_{i} \partial^{m}
\partial_{m} Tr h_{\left (0\right )}= 0\,.
\end{eqnarray}

\newpage
Finally, for the $\left (i j\right )$ part of the EOM

\begin{eqnarray}
\partial^{m} \partial_{m} b_{\left (0\right ) j}^{i} &=& 2 b_{\left (2\right
) j}^{i}  \label{b0EOMfinal} \\
\left(D^{2} h_{\left (0\right )}\right)_{j}^{i} &=& 2 g_{\left (2\right )
j}^{i} + 2 \delta _{j}^{i} Tr g_{\left (2\right )} - 8 b_{\left (2\right )
j}^{i}  \label{h0EOMfinal}
\end{eqnarray}

where

\begin{equation}
\left (D^{2} g_{\left (n\right )}\right )_{i j} = \partial_{i} \partial_{j}
\left(Tr g_{\left(n\right)}\right) + \partial^{m} \partial_{m}
g_{\left(n\right)ij} - \left(\partial_{i} \partial_{k} g_{\left(n\right)
j}^{k} + \partial_{j} \partial_{k} g_{\left (n\right) i}^{k} \right)
\end{equation}

is the general form of the $D^{2}$ operator, introduced in Ref.\cite%
{Johansson:2012fs}, because the covariant derivatives are with respect to the
background Minkowski metric $\eta _{ij}$.

Evaluating the AdS$_{4}$ metric (\ref{linGauss}),(\ref{FGperturbation}), the
different parts of Eq.(\ref{totvaraction}) can be expanded in the following form.
Initially the determinant and the variation of the metric give

\begin{eqnarray}
\sqrt{-h} &=& \frac{\sqrt{-\tilde{g}}}{\rho^{3/2}} = \frac{1}{\rho^{3/2}}
\left(1 + \frac{1}{2} Trc \right) \,, \\
\left(h^{-1} \delta h \right)_{j}^{i} &=& \left(\tilde{g}^{-1} \delta
\tilde{g} \right)_{j}^{i} = \left(\eta^{im} - c^{im}\right) \delta c_{mj} \,,
\end{eqnarray}

while the coefficient of $\delta$K reads

\begin{equation}
\delta _{\left [k \ell \right]}^{\left [ij \right]} G_{j}^{\ell} =
- 3 b_{\left (0\right )k}^{i} - 3 \rho b_{\left (2\right )k}^{i} + 3 \rho
^{3/2}b_{\left (3\right )k}^{i} \,.
\end{equation}

The remaining terms are the coefficients of the variation of the metric
and adopt the form

\begin{eqnarray}
\delta_{\left [k \ell \right ]}^{\left[i j \right]} K_{i}^{m}
G_{j}^{\ell} &=& -3 b_{\left (0\right )k}^{m} + 3 \rho \left[ 2 \left(b_{\left
(0\right )} b_{\left (2\right )}\right)_{k}^{m} - b_{\left (2\right )k}^{m}
+ \left(b_{\left (0\right )} g_{\left (2\right )}\right)_{k}^{m} +
\left(b_{\left (0\right )} b_{\left (2\right )}\right)_{k}^{m} \log \rho %
\right]  \nonumber \\
&+& 3 \rho^{3/2} \left[ b_{\left (3\right )k}^{m} + \frac{3}{2}
\left(b_{\left (0\right )} g_{\left (3\right )}\right)_{k}^{m} + \frac{3}{2}
\left(b_{\left (0\right )} b_{\left (3\right )}\right)_{k}^{m} \log \rho %
\right] \,, \nonumber
\end{eqnarray}

and

\begin{eqnarray}
& & \delta_{\left [k \ell \right ]}^{\left [i j \right ]} \frac{1}{N}
\left( \nabla _{\rho } G_{i}^{k} - \nabla_{i} G_{\rho }^{k} \right) \nonumber \\ 
& & = 3 b_{\left (0\right )\ell}^{j} + 3 \rho \left[ -2 \left(b_{\left (0\right )}
b_{\left (2\right )}\right)_{\ell}^{j} - b_{\left (2\right )\ell}^{j} -
\left(b_{\left (0\right )} g_{\left (2\right )}\right)_{\ell}^{j}
+ \delta_{\ell}^{j} \left(2 Trb_{\left (0\right )}
b_{\left(2\right )} + \right. \right. \nonumber \\
& &  \left. \left.  + Trb_{\left (0\right )}g_{\left (2\right )} +
Trb_{\left (0\right )} b_{\left(2\right )}\log \rho \right) -
\left(b_{\left (0\right )} b_{\left (2\right )}\right)_{\ell}^{j} \log \rho
\right] + 3 \rho^{3/2} \left[ 2 b_{\left (3\right )\ell}^{j} + \right. \nonumber \\
& & \left. + \frac{3}{2} \delta_{\ell}^{j} Trb_{\left (0\right )} g_{\left (3\right )}
- \frac{3}{2} \left(b_{\left (0\right )} g_{\left (3\right )}\right)_{\ell}^{j} - \frac{3}{%
2} \left(b_{\left (0\right )} b_{\left (3\right )}\right)_{\ell}^{j} \log \rho
+ \frac{3}{2} Trb_{\left (0\right )} b_{\left (3\right )} \delta
_{\ell}^{j} \log \rho \right] \,. \nonumber
\end{eqnarray}

The third term in the second line of (\ref{totvaraction}) vanishes in the
linearized case. Here, the terms $b_{\left (0\right )} b_{\left(2\right
)},b_{\left (0\right )} g_{\left(2\right )},b_{\left (0\right )}
b_{\left(3\right )}$ and $b_{\left (0\right )} g_{\left(3\right )}$ are of
order $\mathcal{O} \left(c^2 \right)$. Demanding an action up to quadratic
order in $c_{ij}$, Eq. (\ref{totvaraction}) adopts the form

\small
\begin{equation}
\delta I_{critical} = \frac{1}{32 \pi G} \int \limits _{\partial
\mathcal{M}} d^{3}x \left( 6 \rho^{-3/2} b_{\left(0\right)ij} \delta
b_{\left(0\right)}^{ij} - 6 b_{\left(3\right)ij} \delta b_{\left(0\right
)}^{ij} + 9 b_{\left(3\right )ij} \log \rho \delta b_{\left (0\right )}^{ij}
+ 9 b_{\left(3\right)ij} \delta h_{\left (0\right)}^{ij} \right) \,.
\label{linvaractiondiv}
\end{equation}

\normalsize

Despite the fact that the $b_{\left(2\right )ij}$ contribution is divergent,
the field equations (\ref{b0EOMfinal}) show that it is a total derivative,
so that it can be dropped. From the above derivation, it is evident that the
variation of the action is not finite, due to the presence of a logarithmic
term. This actually corresponds to a divergent logarithmic stress-energy
tensor, as it the conjugate of the source $b_{\left(0\right)ij}$. In turn,
the holographic response to the Einstein source $%
h_{\left(0\right)ij}$ is finite. Following standard holographic
renormalization and the formulation of Refs. \cite%
{Skenderis:2009nt,Alishahiha:2010bw}, we track these divergences at the
level of the action.

Using the EOM (\ref{EOM}), the Eq.(\ref{CGonshell}) can be written as

\begin{equation}
I_{critical} = - \frac{1}{32 \pi G} \int \limits_{\mathcal{M}} d^{4}x \sqrt{%
-g} G_{\nu}^{\mu} G_{\mu}^{\nu} \,.
\label{CGonshell2}
\end{equation}

After some algebraic manipulation and taking into account the linearized
EOM, the square of the linearized Einstein tensor reads

\begin{equation}
G_{\nu}^{\mu} G_{\mu}^{\nu} = 9 Trb_{\left(0\right)}^{2} + 18 \rho
Trb_{\left(0\right)}b_{\left(2\right)} - 18 \rho^{3/2}
Trb_{\left(0\right)}b_{\left(3\right)} \,.
\end{equation}

Putting a cutoff scale at radius $\rho = \varepsilon$, the action (\ref%
{CGonshell2}) can be cast in the form,

\begin{eqnarray}
I_{critical} &=& - \frac{1}{64 \pi G} \int
d^{3}x \int _{\rho =\varepsilon } d\rho \frac{\sqrt{ -\tilde{g}}}{\rho ^{3/2
+1}} G_{\nu}^{\mu} G_{\mu}^{\nu}  \notag \\
&=& - \frac{9}{64\pi G} \int d^{3}x \int _{\rho = \varepsilon } d\rho
\frac{\sqrt{ -\tilde{g}}}{\rho ^{3/2+1}} \left( Trb_{\left(0\right)}^{2} + 2
\rho Trb_{\left(0\right)}b_{\left(2\right)} - 2 \rho^{3/2}
Trb_{\left(0\right)}b_{\left(3\right)} \right)  \notag \\
&=& \frac{9}{32 \pi G} \int \limits _{\partial \mathcal{M}} d^{3}x \left(
Trb_{\left(0\right)} b_{\left(3\right)} \log \varepsilon + \frac{1}{3}
\varepsilon^{-3/2} Trb_{\left(0\right)}^{2} + 2 \varepsilon^{-1/2}
Trb_{\left(0\right)} b_{\left(2\right)} \right) \,.
\end{eqnarray}

\subsection{Counterterms}
All terms tend to infinity at the conformal boundary ($\varepsilon = 0$).
These divergences generate the infinities previously seen at the variation
of the action (\ref{linvaractiondiv}). In order to render the action finite,
proper counterterms have to be added. In the first place we invert the
series as follows

\begin{equation}
b_{\left (0\right )ij} = \rho \partial _{\rho }c_{ij} - \rho \left (b_{\left
(2\right )ij} + g_{\left (2\right )ij} + b_{\left (2\right )ij}\log \rho
\right ) - \rho ^{3/2}\left (g_{\left (3\right )ij} + b_{\left (3\right
)ij}\log \rho \right ) \,.  \label{invertser1}
\end{equation}

The combination

\begin{eqnarray}
\frac{1}{3} \rho^{1/2} \partial_{\rho } c_{ij} \partial_{\rho } c^{ij} &=&
\frac{2}{3} Trb_{\left (0\right )} b_{\left (3\right )} + Trb_{\left
(0\right )} g_{\left (3\right )} + Trb_{\left (0\right )} b_{\left (3\right
)} \log \rho + \frac{1}{3} \rho^{-3/2} Trb_{\left (0\right )}^{2}  \notag \\
&+& \frac{2}{3} \rho^{-1/2} \left( Trb_{\left (0\right )} b_{\left (2\right
)} + Trb_{\left (0\right )} g_{\left (2\right )} + Trb_{\left (0\right )}
b_{\left (2\right )} \log \rho \right) \,,  \label{counter1}
\end{eqnarray}

cancels the leading order logarithmic divergence of the action but
introduces new infinities plus a finite contribution. Taking into account
the Eqs. (\ref{b0EOMfinal},\ref{h0EOMfinal}), the following term can also be
written as

\begin{eqnarray}
Trb_{\left(0\right)} g_{\left(2\right)} = b_{\left(0\right)}^{ij}
g_{\left(2\right)ij} &=& \frac{1}{2} b_{\left(0\right)}^{ij}
\left(D^{2}h_{\left(0\right)}\right)_{ij} - Trb_{\left(0\right)}
Trg_{\left(2\right)} + 4 b_{\left(0\right)}^{ij} b_{\left(2\right)ij}  \notag
\\
&=& \frac{1}{2} b_{\left(0\right)}^{ij}
\left(D^{2}h_{\left(0\right)}\right)_{ij} + 4 Trb_{\left(0\right)}
b_{\left(2\right )}  \notag \\
&=& \frac{1}{2} h_{\left(0\right)}^{ij} \partial^{m} \partial_{m}
b_{\left(0\right)ij} + 4 Trb_{\left(0\right)} b_{\left(2\right)}  \notag \\
&=& h_{\left(0\right)}^{ij} b_{\left(2\right)ij} + 4 Trb_{\left(0\right)}
b_{\left(2\right)}  \notag \\
&=& Trh_{\left(0\right)} b_{\left(2\right)} + 4 Trb_{\left(0\right)}
b_{\left(2\right)} \,,  \label{counter2}
\end{eqnarray}

where integration by parts was performed passing from the second to the
third line. Moreover, inverting the series, one produces the expressions

\begin{eqnarray}
c^{ij} \partial^{m} \partial_{m} \partial_{\rho} c_{ij} &=& 2 \rho^{-1}
\left(Trh_{\left(0\right)} b_{\left(2\right)} + Trb_{\left(0\right)}
b_{\left(2\right)} \log \rho \right) + \mathcal{O} \left(\rho^{0}\right) \,,
\label{counter3} \\
\partial_{\rho} c^{ij} \partial^{m} \partial_{m} \partial_{\rho } c_{ij} &=&
\rho ^{-2} b_{\left(0\right)}^{ij} \partial^{m} \partial_{m}
b_{\left(0\right)ij} + \mathcal{O} \left(\rho^{-1}\right) = 2 \rho^{-2}
Trb_{\left(0\right)} b_{\left(2\right)} + \mathcal{O} \left(\rho^{-1}
\right) \,.  \label{counter4}
\end{eqnarray}

There is a linear combination of the terms in Eqs. (\ref{counter1}) - (\ref%
{counter4}), that cancels the divergences up to finite terms. More
specifically, this can be written as

\begin{eqnarray}
& & \frac{1}{3} \rho^{1/2} \left( \partial_{\rho} c_{ij} \partial_{\rho}
c^{ij} - c^{ij} \partial^{m} \partial_{m} \partial_{\rho} c_{ij} - 2 \rho
\partial_{\rho} c^{ij} \partial^{m} \partial_{m} \partial_{\rho } c_{ij}
\right) =  \notag \\
&=& Trb_{\left(0\right)} b_{\left(3\right)} \log \rho + 2 \rho^{-1/2}
Trb_{\left(0\right)} b_{\left(2\right)} + \frac{1}{3} \rho^{-3/2}
Trb_{\left(0\right)}^{2} + \frac{2}{3} Trb_{\left(0\right)}
b_{\left(3\right)} + Trb_{\left(0\right)} g_{\left(3\right)} \,.  \notag
\end{eqnarray}

Hence, the counterterm action obtains the form

\begin{equation}
I_{ct} = - \frac{3}{32\pi G} \int \limits_{\partial \mathcal{M}} d^{3}x
\rho^{1/2} \left( \partial_{\rho }c_{ij} \partial_{\rho }c^{ij} - c^{ij}
\partial^{m} \partial_{m} \partial_{\rho }c_{ij} - 2 \rho \partial_{\rho
}c^{ij} \partial^{m} \partial_{m} \partial_{\rho } c_{ij} \right) \,.
\label{countertermlin}
\end{equation}

This expression can be covariantized after performing the proper rescaling
of the metric and its perturbation. The respective metric field can be
written as $h_{ij} = \left( \eta_{ij} + c_{ij} \right)/ \rho$. The extrinsic curvature
obtains the form $K_{ij} = \frac{1}{\rho} \eta_{j}^{i} - \kappa_{ij}$ where
$\kappa_{ij} = \partial_{\rho} c_{ij}$. Hence, the fully covariant form of the
counterterms can be cast in the following form

\begin{equation}
I_{ct} = \frac{3}{32 \pi G} \int \limits_{\partial \mathcal{M}} d^{3}x
\sqrt{-h} \left( 2K - K_{ij} K^{ij} - 3 + \frac{1}{N} K^{ij} D^{m} D_{m} K_{ij}
- \frac{1}{2N} D^{m} D_{m} K \right) \,.
\label{countertermlcovariant}
\end{equation}

Our renormalized AdS action relies on the addition of extrinsic counterterms
on top of a bulk topological invariant. This, in principle, provides a different
starting point from the one proposed in Ref.\cite{Johansson:2012fs}. The difference
stems from the use, in the latter reference, of a Dirichlet boundary conditions
for the metric $h_{ij}$.

The variation of the counterterm gives

\begin{eqnarray}
\delta I_{ct} &=& - \frac{3}{32 \pi G} \int \limits _{\partial \mathcal{M}}
d^{3}x \rho^{1/2} \left[ - \partial^{m} \partial_{m} \partial_{\rho }c_{ij}
\delta c^{ij} + \left( 2 \partial_{\rho }c_{ij} - \partial_{m} \partial^{m}
c_{ij} - 4 \rho \partial^{m} \partial_{m} \partial_{\rho} c_{ij} \right)
\delta \left(\partial_{\rho}c^{ij}\right) \right]  \notag \\
&=& - \frac{3}{16 \pi G} \int \limits _{\partial \mathcal{M}} d^{3}x
\rho^{1/2} \partial_{\rho }c_{ij} \delta \left(\partial_{\rho}c^{ij}\right) \,,
\label{varcountertermslin}
\end{eqnarray}

where the rest of the terms have been dropped as total derivatives.

Thus, evaluating Eq. (\ref{varcountertermslin}) and adding on top of Eq. (\ref%
{linvaractiondiv}), the variation of the total action $I_{tot} =
I_{critical} + I_{ct}$ reads

\begin{eqnarray}
\delta I_{total} &=& \frac{1}{32 \pi G} \int \limits_{\partial \mathcal{M}}
d^{3}x \left(6 \rho^{-3/2} b_{\left(0\right)ij} \delta
b_{\left(0\right)}^{ij} - 6 b_{\left(3\right)ij} \delta
b_{\left(0\right)}^{ij} + 9 b_{\left(3\right)ij} \log \rho \delta b_{\left
(0\right)}^{ij} + 9 b_{\left(3\right)ij} \delta h_{\left(0\right)}^{ij}
\right.  \notag \\
& & \left. - 6 \rho^{-3/2} b_{\left(0\right)ij} \delta
b_{\left(0\right)}^{ij} - 6 b_{\left(3\right)ij} \delta b_{\left(0\right
)}^{ij} - 9 b_{\left(3\right)ij} \log \rho \delta b_{\left(0\right)}^{ij} -
9 g_{\left(3\right)ij} \delta b_{\left(0\right)}^{ij} \right)  \notag \\
&=& \frac{1}{32 \pi G} \int \limits_{\partial \mathcal{M}} d^{3}x \left(- 12
b_{\left(3\right)ij} \delta b_{\left(0\right)}^{ij} - 9 g_{\left(3\right)ij}
\delta b_{\left(0\right)}^{ij} + 9 b_{\left(3\right)ij} \delta
h_{\left(0\right)}^{ij} \right) \,.
\end{eqnarray}

\subsection{Holographic correlation functions}

The functional derivatives of the sources are finite giving rise to
holographic energy-momentum tensors. Hence, the one-point functions
around a flat background are given by

\begin{equation}
\langle T_{ij} \rangle = 2 \frac{\delta I_{total}}{\delta
h_{\left(0\right)}^{ij}} = \frac{9}{16 \pi G} b_{\left(3\right)ij} \,,
\label{standardstressenergy}
\end{equation}

what is the holographic dual to the Einstein source $h_{\left(0\right)ij}$, and

\begin{equation}
\langle t_{ij} \rangle = 2 \frac{\delta I_{total}}{\delta
b_{\left(0\right)}^{ij}} = - \frac{3}{16 \pi G} \left( 4 b_{\left(3\right)ij}
+ 3 g_{\left(3\right)ij} \right) \,,
\label{logonepoint}
\end{equation}

is the dual to the logarithmic source $b_{\left(0\right)ij}$.

Following the AdS/CFT dictionary, the variation of these
correlators with respect to the sources provide the two-point
correlation functions. As a result, they read

\begin{eqnarray}
\langle T_{ij} \left(x\right) T_{kl} \left(x'\right) \rangle &=&
- 2i \frac{\delta}{\delta h_{\left(0\right)}^{kl} \left(x'\right)}
\langle T_{ij} \left(x\right) \rangle = -\frac{9i}{8 \pi G}
\frac{\delta b_{\left(3\right)ij} \left(x\right)}{\delta h_{\left(0\right)}^{kl} \left(x'\right)} = 0
\label{TTcorrelator}\\
\langle T_{ij} \left(x\right) t_{kl} \left(x'\right) \rangle &=&
- 2i \frac{\delta}{\delta b_{\left(0\right)}^{kl} \left(x'\right)}
\langle T_{ij} \left(x\right) \rangle = - 2i \frac{\delta}{\delta h_{\left(0\right)}^{kl} \left(x'\right)}
\langle t_{ij} \left(x\right) \rangle \nonumber \\
&=& -\frac{9i}{8 \pi G} \frac{\delta b_{\left(3\right)ij} \left(x\right)}{\delta b_{\left(0\right)}^{kl} \left(x'\right)} =
\frac{9i}{8 \pi G} \frac{\delta g_{\left(3\right)ij} \left(x\right)}{\delta h_{\left(0\right)}^{kl} \left(x'\right)}  \label{Ttcorrelator}\\
\langle t_{ij} \left(x\right) t_{kl} \left(x'\right) \rangle &=&
- 2i \frac{\delta}{\delta b_{\left(0\right)}^{kl} \left(x'\right)}
\langle t_{ij} \left(x\right) \rangle = \frac{3i}{8 \pi G} \left( 4 \frac{\delta b_{\left(3\right)ij} \left(x\right)}{\delta b_{\left(0\right)}^{kl} \left(x'\right)} + 3 \frac{\delta g_{\left(3\right)ij} \left(x\right)}{\delta b_{\left(0\right)}^{kl} \left(x'\right)}\right)
\label{ttcorrelator}
\end{eqnarray}

The i factor in the two-point point functions comes from
the generating functional when written in Lorentzian signature.
More precisely, in the context of AdS/CFT correspondence a relation
between the generating functional and the on-shell action of the
type $W_{L}\sim i I_{L}$. This choice yields the formulas displayed
above \cite{Skenderis:2009nt}.

One can notice that the norm of the stress-energy tensor is zero, as
expected in a LCFT. This is an immediate consequence of the fact that
no Einstein mode can source a logarithmic mode. The latter assertion
was remarked in Ref.\cite{Johansson:2012fs}, where the mode analysis
allows to calculate the aforementioned functional derivatives.

Actually, the EOM (\ref{trb0}-\ref{divb0}) show that $b_{3}$ is a
transverse and traceless mode while $g_{3}$ it is just traceless.
The latter is a consequence of the presence of non-Einstein modes
in the theory, whereas in Einstein Gravity $g_{\left(3\right)ij}$
is both transverse and traceless and determines the holographic
stress-energy tensor. This property leads to the York decompositions
of the $g_{\left(3\right)}$ mode, which reads:

\begin{equation}\label{g3decomp}
g_{\left(3\right)ij} = \nabla_{i} V_{j}^{\left(3\right)} +
\nabla_{j} V_{i}^{\left(3\right)} + g_{\left(3\right)ij}^{TT} +
\left( \nabla_{i} \nabla{j} - \frac{1}{3} \eta_{ij} \nabla^{2} \right) S^{\left(3\right)} \,.
\end{equation}

Each one of the terms contribute independently to different pieces
of the one-point function $t_{ij}$ which now consists of:
i) a transverse vector  $V_{i}$,
ii) a transverse traceless part $t_{ij}^{TT}$, which is the logarithmic conjugate of $T_{ij}$,
iii) and a scalar S.

Consequently, from Eq. (\ref{logonepoint}) we get the following three operators:

\begin{eqnarray}
\langle t_{ij}^{TT} \rangle &=& - \frac{3}{16 \pi G} \left( 4 b_{\left(3\right)ij}
+ 3 g_{\left(3\right)ij}^{TT} \right) \\
\langle V_{i} \rangle &=& - \frac{9}{16 \pi G} V_{i}^{\left(3\right)} \\
\langle S \rangle &=& - \frac{9}{16 \pi G} S^{\left(3\right)} \,.
\end{eqnarray}

These values are justified considering that the vector and the
scalar operator contributions come explicitly from $g_{\left(3\right)ij}$,
while the logarithmic stress-energy tensor $t_{ij}^{TT}$ is sourced by both parts.

Considering that $b_{\left(3\right)ij}$ is transverse and traceless leads
to only one non-vanishing mixed correlator in Eq. (\ref{Ttcorrelator}),
the one between the two stress-energy tensors. Given the mode dependence
on the sources in \cite{Johansson:2012fs}, we obtain that

\begin{equation}
\langle T_{ij} \left(x\right) t_{kl}^{TT} \left(0\right) \rangle =
- \frac{1}{2 \pi^{3}} \frac{3}{2 G} \hat{\Delta}_{ij,kl} \frac{1}{|x^{2}|} \,,
\end{equation}

where

\begin{eqnarray}
\hat{\Delta}_{ij,kl} &=& \frac{1}{2} \left( \hat{\Theta}_{ik} \hat{\Theta}_{jl}
+ \hat{\Theta}_{il} \hat{\Theta}_{jk} - \hat{\Theta}_{ij} \hat{\Theta}_{kl} \right) \\
\hat{\Theta}_{ij} &=& \partial_{i} \partial_{j} - \eta_{ij} \Box \,.
\end{eqnarray}

Finally, from (\ref{ttcorrelator}) we get three different correlators,
each one corresponding to the vector, scalar and transverse traceless
operators. The former ones obtain contributions only from the $g_{\left(3\right)ij}$
functional derivatives. Hence:

\begin{eqnarray}
\langle V_{i} \left(x\right) V_{j} \left(0\right) \rangle &=& \frac{9i}{8 \pi G}
\left( \frac{\delta g_{\left(3\right)ij} \left(x\right)}{\delta b_{\left(0\right)}^{kl} \left(0\right)} \right)_{V}
= \frac{1}{2 \pi^{3}} \frac{9i}{8 \pi G} \int d^{3}p e^{ipx} \left( \frac{\delta g_{\left(3\right)ij}}{\delta b_{\left(0\right)}^{kl}} \left(p\right) \right)_{V} \nonumber \\
&=& - \frac{1}{2 \pi^{3}} \frac{1}{4 G} \hat{\Theta}_{ij} \frac{1}{|x^{2}|}
\end{eqnarray}

and

\begin{eqnarray}
\langle S \left(x\right) S \left(0\right) \rangle &=& \frac{9i}{8 \pi G}
\left( \frac{\delta g_{\left(3\right)ij} \left(x\right)}{\delta b_{\left(0\right)}^{kl} \left(0\right)} \right)_{S}
= \frac{1}{2 \pi^{3}} \frac{9i}{8 \pi G} \int d^{3}p e^{ipx} \left( \frac{\delta g_{\left(3\right)ij}}{\delta b_{\left(0\right)}^{kl}} \left(p\right) \right)_{S} \nonumber \\
&=& \frac{1}{2 \pi^{3}} \frac{3}{8 G} \frac{1}{|x^{2}|} \,.
\end{eqnarray}

The latter two-point function is the one corresponding to the logarithmic
stress-energy tensors. It receives contributions from the transverse traceless
part of both  $b_{\left(3\right)i}$ and $g_{\left(3\right)}$. One may rewrite
Eq. (\ref{ttcorrelator}) as

\begin{equation}
\langle t_{ij}^{TT} \left(x\right) t_{kl}^{TT} \left(x'\right) \rangle =
- \frac{4}{3} \langle T_{ij} \left(x\right) t_{kl}^{TT} \left(x'\right) \rangle
+ \frac{9i}{8 \pi G} \left( \frac{\delta g_{\left(3\right)ij} \left(x\right)}{\delta b_{\left(0\right)}^{kl} \left(x'\right)} \right)_{TT} \,.
\end{equation}

In this case we obtain that

\begin{eqnarray}
\langle t_{ij}^{TT} \left(x\right) t_{kl}^{TT} \left(0\right) \rangle &=&
- \frac{4}{3} \langle T_{ij} \left(x\right) t_{kl}^{TT} \left(0\right) \rangle
+  \frac{1}{2 \pi^{3}} \frac{9i}{8 \pi G} \int d^{3}p e^{ipx} \left( \frac{\delta g_{\left(3\right)ij}}{\delta b_{\left(0\right)}^{kl}} \left(p\right) \right)_{TT} \nonumber \\
&=& - \frac{1}{2 \pi^{3}} \frac{3}{2 G} \hat{\Delta}_{ij,kl} \frac{\log |x^{2}| + C + 4 \gamma - 4/3}{|x^{2}|} \,,
\end{eqnarray}

where C is a numerical constant. In general, the logarithmic stress tensor
is defined up to the addition of a multiple of $\langle T_{ij} \rangle$.
Therefore, taking advantage of this freedom, we redefine $t_{ij}^{TT}$
as $t_{ij}^{TT} \rightarrow - \left(C/4 + \gamma - 1/3 \right)T_{ij}$,
canceling all the numerical constants appearing in the numerator.
Hence, we obtain that

\begin{equation}
\langle t_{ij}^{TT} \left(x\right) t_{kl}^{TT} \left(0\right) \rangle =
- \frac{1}{2 \pi^{3}} \frac{3}{2 G} \hat{\Delta}_{ij,kl} \frac{\log |x^{2}|}{|x^{2}|} \,.
\end{equation}

\section{Conclusions}

In the present paper, we have computed holographic correlation functions
in Einstein-Weyl gravity at the critical point. We have applied holographic
techniques to an equivalent form of the Critical Gravity action, given by
Eq. (\ref{criticalgravity}), where the curvature-squared part are expressed
as the difference between the Weyl$^{2}$ and the GB terms.
The GB term, with its coupling fixed by the above argument, provides partial
renormalization of the variation of the action, such that the divergent
pieces can be attributed to the non-Einstein part in the curvature (Bach
tensor).

In turn, for Einstein modes, both the action and its variation are not only
finite, but identically zero \cite{Miskovic:2014zja,Anastasiou:2016jix}.
The vanishing of the holographic stress tensor for Einstein spacetimes
\cite{Anastasiou:2017rjf}, together with a zero mass
and entropy for Einstein black holes, indicates that Critical Gravity turns
somehow trivial within that sector.

Additional counterterms, which depend on the extrinsic curvature and its
covariant derivatives, are needed when the logarithmic source is switched
on at the boundary. The departure from the Einstein condition by including
log terms in the metric modifies the asymptotic form of the Riemann tensor
at leading order.

The addition of these terms makes the action principle not suitable for
imposing a Dirichlet boundary condition in $h_{ij}$. However, variation
of the action is finite and written down in terms of variations of
$h_{\left(0\right)ij }$ and $b_{\left(0\right)ij }$. In other words,
the counterterms in (\ref{countertermlcovariant}) provide a well-posed
variational principle by fixing the holographic sources on the conformal
boundary. In this sense, our boundary conditions are compatible with the
holographic description
of AdS gravity with relaxed asymptotic behavior.

In Einstein gravity with standard AdS boundary conditions, the fall-off
of the curvature tensor determines the coupling of the GB term in
Eq. (\ref{reneinsteinads}). The locally equivalent boundary term to the
GB invariant is the 2nd Chern form, which is a given polynomial of the
extrinsic and intrinsic curvatures. Intrinsic counterterms presented in
Refs. \cite{Balasubramanian:1999re,Emparan:1999pm} are worked out as a
truncation of the series coming from taking a FG expansion on the extrinsic
curvature \cite{Miskovic:2009bm}.

A similar comparison, this time, between the counterterms in Eq.(\ref{countertermlcovariant})
and the ones presented in Ref.\cite{Johansson:2012fs} might be worked out
adding and substracting the corresponding generalized Gibbons-Hawking term for
Critical Gravity as a higher-derivative theory

\begin{equation}
I_{GGH} = \frac{1}{2 \kappa^{2}} \int \limits _{\partial \mathcal{M}} d^{3}x F^{ij} \left( K \gamma_{ij} - K_{ij} \right)\,.
\end{equation}

On the other hand, and because $b_{(0)}$ is neither a parameter of the
theory nor a covariant field in the Lagrangian, there is no direct way
to fine tune the GB coupling to incorporate the information on the modified
asymptotic curvature. Therefore, it remains as an open problem how to mimic
the effect of Topological Regularization in presence of a log boundary source.

\begin{acknowledgments}
The authors thank O. Miskovic and T. Zojer for interesting discussions. G.A. is a
Universidad Andres Bello (UNAB) Ph.D. Scholarship holder, and his work is supported
by Direcci\'{o}n General de Investigaci\'{o}n (DGI-UNAB). The work of R.O. is funded
in part by FONDECYT Grant No. 1170765, UNAB Grant DI-1336-16/R and CONICYT Grant DPI
20140115.
\end{acknowledgments}

\bibliographystyle{JHEP}

\bibliography{referencescritical}

\end{document}